\documentclass{article}
\usepackage[tight]{subfigure}
\usepackage{spconf,amsmath,graphicx}
\ninept
\usepackage{bm}
\usepackage{amsthm}
\usepackage{amsfonts}
\usepackage{algorithm}
\usepackage{algpseudocode}
\usepackage{epstopdf}

\title{A Range-Doppler-Angle Estimation Method for Passive Bistatic Radar}
\name{\footnotesize{Liangtian Wan$^{\star}$, Member, IEEE, \qquad Xianpeng Wang$^{\sharp,\dagger}$, Member, IEEE  \qquad  Guoan Bi$^{\ddag}$, Senior Member, IEEE}
}

\address{$^{\star}$ \footnotesize{Key Laboratory for Ubiquitous Network and Service Software of Liaoning Province}, \\
\footnotesize{School of Software, Dalian University of Technology, Dalian 116620, China}. \\
$^{\sharp}$ \footnotesize{State Key Laboratory of Marine Resource Utilization in South China Sea, Hainan University, Haikou 570228, China}.\\
$^{\dagger}$ \footnotesize{College of Information Science and Technology, Hainan University, Haikou 570228, China}.\\
$^{\ddag}$ \footnotesize{School of Electrical and Electronic Engineering, Nanyang Technological University, 639798, Singapore}}

\begin{document}

%
\maketitle
\begin{abstract}
In this paper, an effective target detection and localization method is proposed for a passive bistatic radar (PBR) system. The PBR system consists of a commercial FM radio station, which is a non-cooperative illuminator of opportunity (IO),  referred to as the transmitter antenna and multiple surveillance antennas that form an antenna array, e.g., uniform linear array (ULA). Unlike other literatures where the reference signal is received by a directional antenna, here, the reference signal (direct path) is estimated by beamforming method. Then a modified extensive cancellation algorithm (MECA) based on (least squares) LS method is proposed to solve the disturbance cancellation. After cancelling the disturbance, the matched filter (MF) and LS methods are used for range-Doppler estimation of targets, and then the angles of targets are estimated based on beamforming method. The proposed method is suitable for an antenna array. Simulation results are presented to illustrate the superiority of the proposed MECA disturbance cancellation method and parameter estimation method.
\end{abstract}

\vspace{-1em}
\section{Introduction}
Passive sensing has been widely used in many applications, such as radar, underwater acoustics, seismology, etc \cite{griffiths2005passive,baker2005passive,baczyk2011Reconstruction,yocom2011a,artman2011Imaging}. A typical passive bistatic radar (PBR) system exploits a single non-cooperative illuminator of opportunity (IO), referred to as the transmitted antenna. Compared with active sensing, it has the advantage of low hardware system cost, working without interference with existing wireless systems, etc.

In general, the reference antenna is steered towards the transmitter to collect the direct path signal (reference signal), while a surveillance antenna is used to measure a potential target echo (received signal) \cite{griffiths2005passive,baker2005passive,Colone2009multistage}. However, the detection performance of the methods based on matched filter (MF) \cite{Colone2009multistage} would be degraded without considering strong multipath propagations or clutters.  The extensive cancellation algorithm (ECA) method was proposed in \cite{Colone2009multistage,Colone2009space} to cancel the direct path, clutters, and their delay, Doppler shifted versions. Then a least squares (LS) adaptive interference cancellation was proposed in \cite{tan2014space}, which has better cancellation performance than ECA. However, the method proposed in \cite{Colone2009multistage,tan2014space} requires that the length of reference signal is longer than that of the received signal, which is not convenient for batch processing of raw data. In addition, many methods have been proposed to estimate range-Doppler \cite{sun2010applications,hack2014detection,liu2014two,zhang2016joint}, but there are few reports about range-Doppler-angle estimation.

In this paper, a novel range-Doppler-angle estimation method is proposed for PBR system.  The reference signal (direct path) is estimated based on beamforming method. Then, the problem of cancellation of direct path, multipath and clutter signals in PBR system is examined. In real application, the sidelobes of clutters cannot be cancelled completely since the existing disturbance cancellation method can work only when the length of direct path (reference signal) is longer than that of the received array data. In addition, the existing disturbance cancellation method is only suitable for a single antenna. Thus a modified extensive cancellation algorithm (MECA) method base on LS is proposed for disturbance cancellation. After disturbance cancellation, MF and LS methods are used for range-Doppler estimation of target. Finally, the angle of target is estimated based on beamforming method.
\vspace{-1em}
\section{SYSTEM MODEL}
As shown in Fig. 1, the PBR system is equipped with a surveillance array (e.g., uniform linear array (ULA)) to receive the direct path signal (unknown source signal), as well as to acquire the reflected target echo. In addition, the surveillance array can inevitably receive the signals from other sources such as the multipath and clutter echoes reflected or refracted from the ground and surrounding buildings.

Consider the case that the surveillance array forms a ULA with adjacent antennas spaced by $\delta_A$. As mentioned above, the data model of the surveillance array should contain the direct path, multipath/clutter and reflected target echo signals, which can be expressed as
\begin{equation} \label{eq:SS}
\begin{aligned}
x(t,l) = &d(t)e^{j2\pi f_{DA}l} + \sum_{i=1}^{N_C} c_id(t-\tau_{ci})e^{j2\pi f_{CAi}l} \\
&+ \sum_{m=1}^{N_T} a_m d(t-\tau_m)e^{j2\pi f_{TDm}t}e^{j2\pi f_{TAm}l}\\
& + n(t,l), \quad 0\leq t<T_0, 0\leq l<L_A,
\end{aligned}
\end{equation}
where $x(t,l)$ and $n(t,l)$ respectively stand for the received signal and measurement noise of the $l$th antenna at time $t$, $d(t)$ is the direct path signal. $N_C$ and $N_T$ are the number of clutters and targets, respectively. $c_i$ and $a_m$ stand for the complex amplitude of the $i$th clutter and the $m$th target, respectively. $\tau_{ci}$ and $t_m$ stand for the bistatic delay of $i$th clutter and $m$th target, respectively. $f_{DA}$, $f_{CAi}$ and $f_{TAm}$ respectively stand for the angle frequencies of the direct path, the $i$th clutter and the $m$th target, particularly $f_{DA}=\frac{\delta_A}{\lambda} \cos \theta_{DA}$, $f_{CAi}=\frac{\delta_A}{\lambda} \cos \theta_{CAi}$ and $f_{TAm}=\frac{\delta_A}{\lambda} \cos \theta_{TAm}$ with $\lambda$ being the wavelength of the carrier signal. Their corresponding angles are $\theta_{DA}$, $\theta_{CAi}$ and $\theta_{TAm}$, respectively. $f_{TDm}$ is the Doppler frequency of the $m$th target. It should be noted that the clutter does not have Doppler shift. The problem of interest is to estimate the target parameters $\tau_m$, $f_{TDm}$ and $f_{TAm}, m=1,2,\dots,N_T$ from the observation $x(t,l), 0\leq t<T_0, 0\leq l<L_A$.
\begin{figure}
  \centering
  \includegraphics[width=2.5in,height=1in]{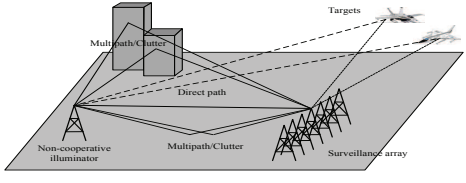}
\caption{A PBR system with a surveillance array and a non-cooperative IO}\label{1}
\end{figure}

\vspace{-1em}
\section{Parameter estimation for PBR system}
\vspace{-1em}
\subsection{Direct path estimation}
First of all, we need to estimate the direct path, which is crucial in the ensuing parameter estimation process. The angle of the direct path can be assumed to be known a priori. This assumption is reasonable since the IO is static. If it is not, it can be estimated by some high accuracy direction-of-arrival (DOA) estimation algorithm such as (multiple signal classification) MUSIC \cite{Schmidt1986Multiple,Krim1996Two}, etc. The direct path can be estimated by using beamforming method, i.e., the ULA is steered to the direction of IO. In particular, given the received array data $\mathbf{X}\in \mathbb{C}^{L_T \times L_A}$, where $L_T$ is the length of a segment of the observation time. The direct path estimation can be expressed as
\begin{equation}\label{DP}
\mathbf{s}_{\mathrm{dp}}=\mathbf{X}\mathbf{a}^*(\theta_{\mathrm{DA}}),
\end{equation}
where $\mathbf{a}(\theta_{\mathrm{DA}})=[1,e^{j2{\pi}f_{\mathrm{DA}}},\cdots ,e^{j2{\pi}(L_A-1)f_{\mathrm{DA}}}]^T$ is the steering vector of the direct path, $\mathbf{s}_{\mathrm{dp}}$ is the direct path estimation.

\vspace{-1em}
\subsection{Disturbance cancellation method}
At present, several cancellation techniques have been proposed in the literature \cite{guner2003direct, colone2006cancellation,cardinali2007comparison}. We will revisit the ECA method by using an LS method \cite{colone2006cancellation}. For the ECA method, a dictionary is constructed, and each column of the matrix, which corresponds to a potential source of disturbance, is a delay and Doppler shifted version of the direct path. The dictionary can be expressed as
\begin{equation}\label{Dic}
\mathbf{Y}=\mathbf{B}[\mathbf{\Lambda} _{-P}\mathbf{S}_{\mathrm{ref}}\cdots\mathbf{\Lambda} _{-1}\mathbf{S}_{\mathrm{ref}}\mathbf{S}_{\mathrm{ref}}\mathbf{\Lambda} _{1}\mathbf{S}_{\mathrm{ref}}\cdots\mathbf{\Lambda} _{-P}\mathbf{S}_{\mathrm{ref}}],
\end{equation}
where $\mathbf{S}_{\mathrm{ref}}=[\mathbf{s}_{\mathrm{dp}}{\kern 2pt}\mathbf{Ds}_{\mathrm{dp}}{\kern 2pt}\mathbf{D}^2\mathbf{s}_{\mathrm{dp}}\cdots \mathbf{D}^{Q-1}\mathbf{s}_{\mathrm{dp}}]$, $\mathbf{D}$ is a matrix that applies a delay of the direct path. $\mathbf{\Lambda} _{p}$ is a diagonal matrix that applies the
phase shift corresponding to the $p$th Doppler bin and  $\mathbf{B}$ is an incidence matrix that selects only the last $N$ rows of the following matrix. The detail of the definition of $\mathbf{D}$,$\mathbf{\Lambda} _{p}$ and $\mathbf{B}$ can be found in \cite{Colone2009multistage}. The columns of matrix $\mathbf{Y}$ define a basis for the $M$-dimensional clutter subspace, where $M=(2P+1)Q$. To minimize residual signal power after disturbance cancellation based on LS error criterion, the cost function can be written as
\begin{equation}\label{LS}
\mathop {\min }\limits_{\mathbf{W}} \left \| \mathbf{X}-\mathbf{YW} \right \|^2.
\end{equation}
The optimized LS solution of the weighting vector can be expressed as $\mathbf{W}=(\mathbf{Y}^H\mathbf{Y})^{-1}\mathbf{Y}^H\mathbf{X}$. Then the output of surveillance array after cancellation can be expressed as
\begin{equation}\label{OSA}
\mathbf{X}_{\mathrm{ECA}}=\mathbf{X}-\mathbf{YW}=\mathbf{PX},
\end{equation}
where $\mathbf{P}=\mathbf{I}_N-\mathbf{Y}(\mathbf{Y}^H\mathbf{Y})^{-1}\mathbf{Y}^H$ is the projection matrix which projects the received matrix $\mathbf{X}$ onto the orthogonal subspace of the disturbance subspace.

\vspace{-1em}
\subsection{A Modified Disturbance Cancellation Method}

It should be noted that the length of direct path $\mathbf{s}_{\mathrm{dp}}$ (reference signal) is longer than that of received array data  in the ECA method proposed in \cite{Colone2009multistage,colone2006cancellation}. In this paper, the estimated direct path $\mathbf{s}_{\mathrm{dp}}=[s_{\mathrm{dp}}[1]\cdots s_{\mathrm{dp}}[T]]$ is regarded as the reference signal. Thus not all the received array data can be used for disturbance cancellation. The first $R-1$ array data samples should be discarded, thus the dimension of the effective array data is $(L_T-R+1) \times L_A$. $R-1$ is the number of additional reference signal samples \cite{Colone2009multistage}. However, the different lengths could cause a problem in real data processing. In general, the time window for parameter estimation is an integer, and a common setting is $1$s \cite{howland2008passive}. This is convenient for both the batch processing of raw data and the comparison between the estimated parameters and the ground truth received by the automatic dependent surveillance broadcast (ADS-B) \cite{adsb} in the ensuing parameter estimation. When disturbance cancellation method mentioned above is used, one condition is that the duration of reference signal (direct path) is larger than $1$s (e.g., $1.01$s). This means we have to process the raw data by $1.01$s once a time. If we collect $500$s real data for testing our proposed algorithm, the time of the raw data cannot match that of ADS-B after processing dozens of times because of the time difference $0.01$s. It is hard to compare the estimated parameters with the ground truth, thus we can hardly know if our method is efficient. The other condition is that the duration  of reference signal equals to $1$s. This means that the duration of the received array data must be shorter than 1s (e.g., $0.99$s), which leads to the fact that the bistatic Doppler is not an integer (The resolution of bistatic Doppler is 1/0.99). When the disturbance cancellation method is used, the bistatic Doppler bins that we want to cancel are integers, thus some Doppler bins at this resolution may not be cancelled completely. This may lead to performance degradation of disturbance cancellation.

Thus a novel disturbance cancellation method is proposed and the length of direct path $\mathbf{s}_{\mathrm{dp}}$ and received array data are identical, which means that the delay and Doppler shifted version of the direct path corresponding to the clutters can be cancelled completely when the duration of reference signal equals to $1$s. According to matrix of equation (11) in \cite{tan2014space}, it can be known that the length of direct path $\mathbf{s}_{\mathrm{dp}}$ and received array data are still different. However, it gives us an inspiration that we can design a similar matrix as mentioned above. We redefine a $L_T\times \overline{Q}$ direct path matrix, where each column is a unique circle shift delay copy of the direct path signal, as

\begin{equation}\label{DPM}
\mathbf{S}_{\mathrm{mref}}=\begin{bmatrix}
s_{\mathrm{dp}}[1] & s_{\mathrm{dp}}[T_0] &\cdots   & s_{\mathrm{dp}}[T_0-\overline{Q}+2]\\
s_{\mathrm{dp}}[2] & s_{\mathrm{dp}}[1] & \cdots  & s_{\mathrm{dp}}[T_0-\overline{Q}+3]\\
\vdots  & \vdots  & \ddots  & \vdots \\
s_{\mathrm{dp}}[T_0] & s_{\mathrm{dp}}[T_0-1] & \cdots  & s_{\mathrm{dp}}[T_0-\overline{Q}+1]
\end{bmatrix}
\end{equation}
where $\overline{Q}$ is the number of cancellation weight for the finite impulse response (FIR) filter. The other process is the same as subsection 3.2. It can be seen that the dimension of $\mathbf{S}_{\mathrm{mref}}$ is identical with $\mathbf{S}_{\mathrm{ref}}$, which means the computation and memory load of this modified ECA (MECA) method are the same as those of ECA method without the issues pointed out at the beginning of this part.

\vspace{-1em}
\section{The Proposed Parameter Estimation method}
For range-Doppler estimation, the most common method is MF, which evaluates the cross-correlation function (2D-CCF) between the received array data $\mathbf{X}$ and the direct path estimation $\mathbf{s}_{\mathrm{dp}}$. In real application, the range-Doppler can be estimated based on only a single antenna and the direct path estimation. Since multiple antennas can be used, the multiple results of 2D-CCF can be summed together to improve the signal-to-noise ratio (SNR), which can improve the detection performance of the targets. The sum of multiple delay-Doppler 2D-CCFs can be expressed as
\begin{equation}\label{2DCCF}
\xi(\tau,f_D)=\sum_{l=1}^{L_A}\sum_{t=1}^{L_T}\mathbf{X}_{\mathrm{MECA}}[l,t]s_{\mathrm{dp}}^*[t-\tau]e^{-j2{\pi}f_Dt/L_T},
\end{equation}
where $\mathbf{X}_{\mathrm{MECA}}$ is the output of surveillance array after doing MECA method, which contains the target echoes. It can be seen from (\ref{2DCCF}) that, if only a single echo is contained in $\mathbf{X}_{\mathrm{MECA}}$, then the 2D-CCF would achieve the maximum at $(\tau, f_D)$. However, if there are several targets contained in $\mathbf{X}_{\mathrm{MECA}}$, the sidelobes of strong target may mask the weak targets due to the self-ambiguity characteristic of FM signals. Moreover, if some strong clutter echoes are not cancelled completely but remain in $\mathbf{X}_{\mathrm{MECA}}$, the sidelobes of clutters would mask the targets of interest. Numerical simulations will be provided to show this phenomena in the next section.

For the problem that sidelobes of strong targets may mask the weak targets, a sequential range-Doppler estimation for the targets of interest is proposed. Since the range-Doppler of strong targets can be detected after MF, the strong targets can be cancelled in the range-Doppler (RD) map which corresponds to the 2D-CCF. Then the weak target masked by the sidelobes of the strong targets can be detected. The same process can be applied until all the targets of interest are detected. Assume that the range-Doppler of the strong targets that have already been detected are $(r_k, f_{Dk}),k=1,\cdots N_{\mathrm{st}}$, where $N_{\mathrm{st}}$ denotes the number of strong targets. Then we can reconstruct a direct path matrix containing the strong targets as follows
\begin{equation}\label{RDPM}
\mathbf{S}_{\mathrm{st}}[k]=\begin{bmatrix}
s_{\mathrm{dp}}[1] & s_{\mathrm{dp}}[T_0]&\cdots   & s_{\mathrm{dp}}[T_0-r_k-r_0+1]\\
s_{\mathrm{dp}}[2] &s_{\mathrm{dp}}[1] &  \cdots  & s_{\mathrm{dp}}[T_0-r_k-r_0+2]\\
\vdots  &\vdots& \ddots  & \vdots \\
s_{\mathrm{dp}}[T_0] & s_{\mathrm{dp}}[T_0-1]&\cdots  & s_{\mathrm{dp}}[T_0-r_k-r_0]
\end{bmatrix},
\end{equation}
$k=1,\cdots N_{\mathrm{st}}$ (It is to be noted here that we will omit the index $k$ in the following for notational convenience), where $r_0$ is a small integer, whose goal is that the strong target can be cancelled in the range bin completely.  The dictionary containing the strong targets can be constructed as
\begin{equation}\label{Dics}
\mathbf{Y}_s=[\mathbf{\Lambda} _{f_{Dk}-f_0}\mathbf{S}_{\mathrm{st}}\cdots\mathbf{\Lambda} _{f_{Dk}}\mathbf{S}_{\mathrm{st}}\cdots\mathbf{\Lambda} _{f_{Dk}+f_0}\mathbf{S}_{\mathrm{st}}],
\end{equation}
where $f_0$ is a small integer, whose goal is that the strong target can be cancelled in the Doppler bin completely. In general, both $r_0$ and $f_0$ can be chosen as 3. Then the output of surveillance array containing the weak targets can be expressed as $\mathbf{X}_{\mathrm{W}}=\mathbf{P}_s{\mathbf{X}}_{\mathrm{MECA}}$, where $\mathbf{P}_s=\mathbf{I}_{L_T}-\mathbf{Y}_s(\mathbf{Y}^H_s\mathbf{Y})^{-1}_s\mathbf{Y}^H_s$ is the projection matrix which projects $\mathbf{X}_{\mathrm{MECA}}$ onto the orthogonal subspace of the subspace of strong targets. Then $\mathbf{X}_{\mathrm{W}}$ is used to do the MF method based on (\ref{2DCCF}) until all the weak targets are detected.

\emph{Remark} 1: It should be noted that both range and Doppler have the resolution limitation. In general,  $r_k$, $k=1,\cdots N_{\mathrm{st}}$ cannot be directly used to form the direct path matrix, as well as the dictionary containing strong targets by using $f_{Dk}$. For example, if the sampling rate is $250$MHz, the range resolution is $3\times 10^8/250\times 10^6=1.2$Km. Thus the $r_k$ used for forming the direct path matrix should be revised as $r_k/1.2$. If the time window is fixed at $1$s, the Doppler resolution is 1Hz.  If the time window is fixed at $0.5$s, the Doppler resolution is 2Hz. At this time, the Doppler used for constructing the dictionary should be revised as $(\cdots f_{Dk}-2,f_{Dk},f_{Dk}+2,\cdots )$.

After the range-Doppler estimation, the remaining task is angle estimation for targets of interest. The value corresponding to the target in the RDmap for each antenna contains the angle information of the target. The values corresponding to multiple antennas can be written as follows
\begin{equation}\label{An}
\mathbf{z}_m=[z_{1m},z_{2m} \cdots z_{L_Am}]^T,m=1,\cdots, N_T.
\end{equation}

\begin{figure}
  \subfigure[]{
    \includegraphics[width=1.6in,height=1in]{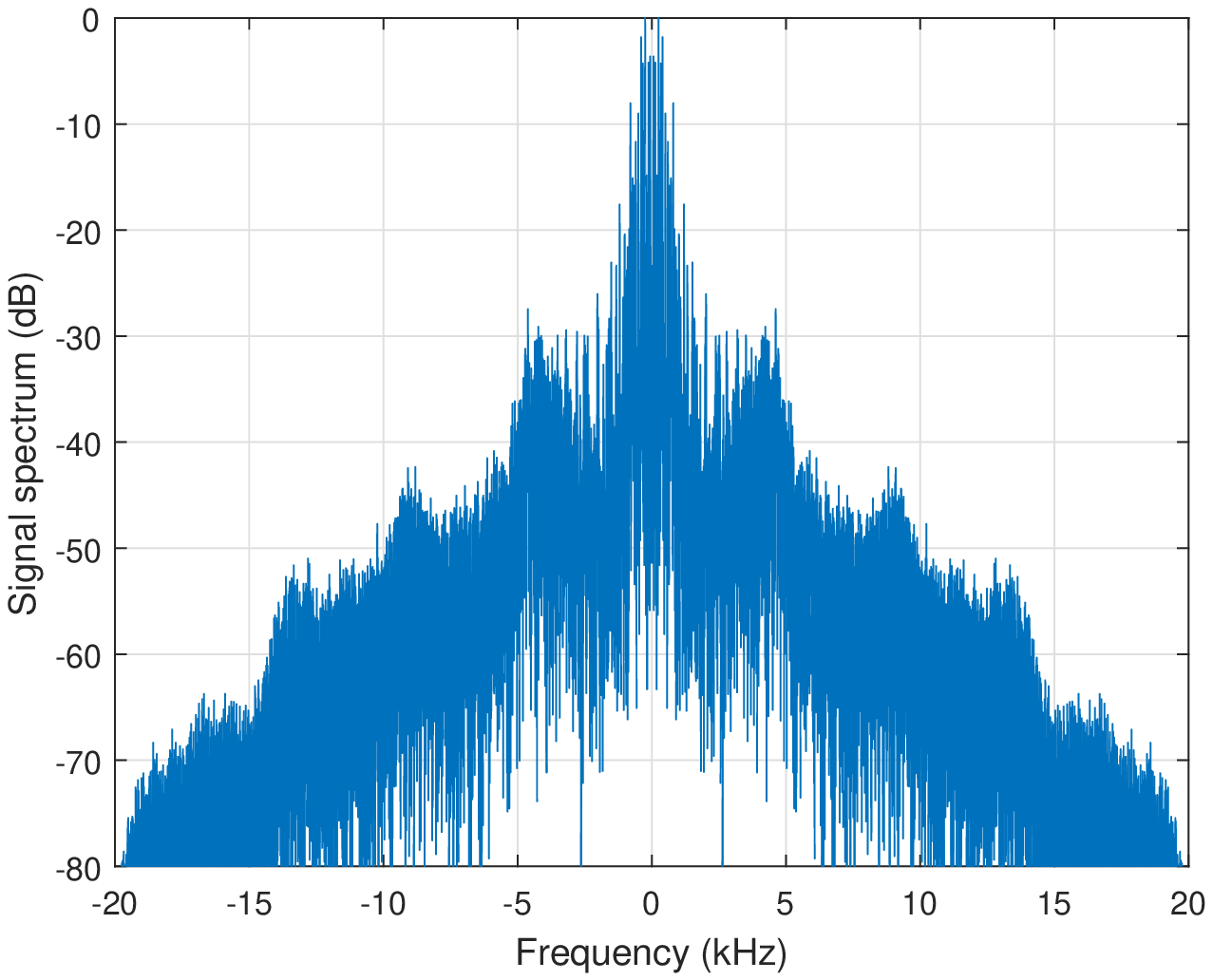}} %
  \subfigure[]{
    \includegraphics[width=1.56in,height=1in]{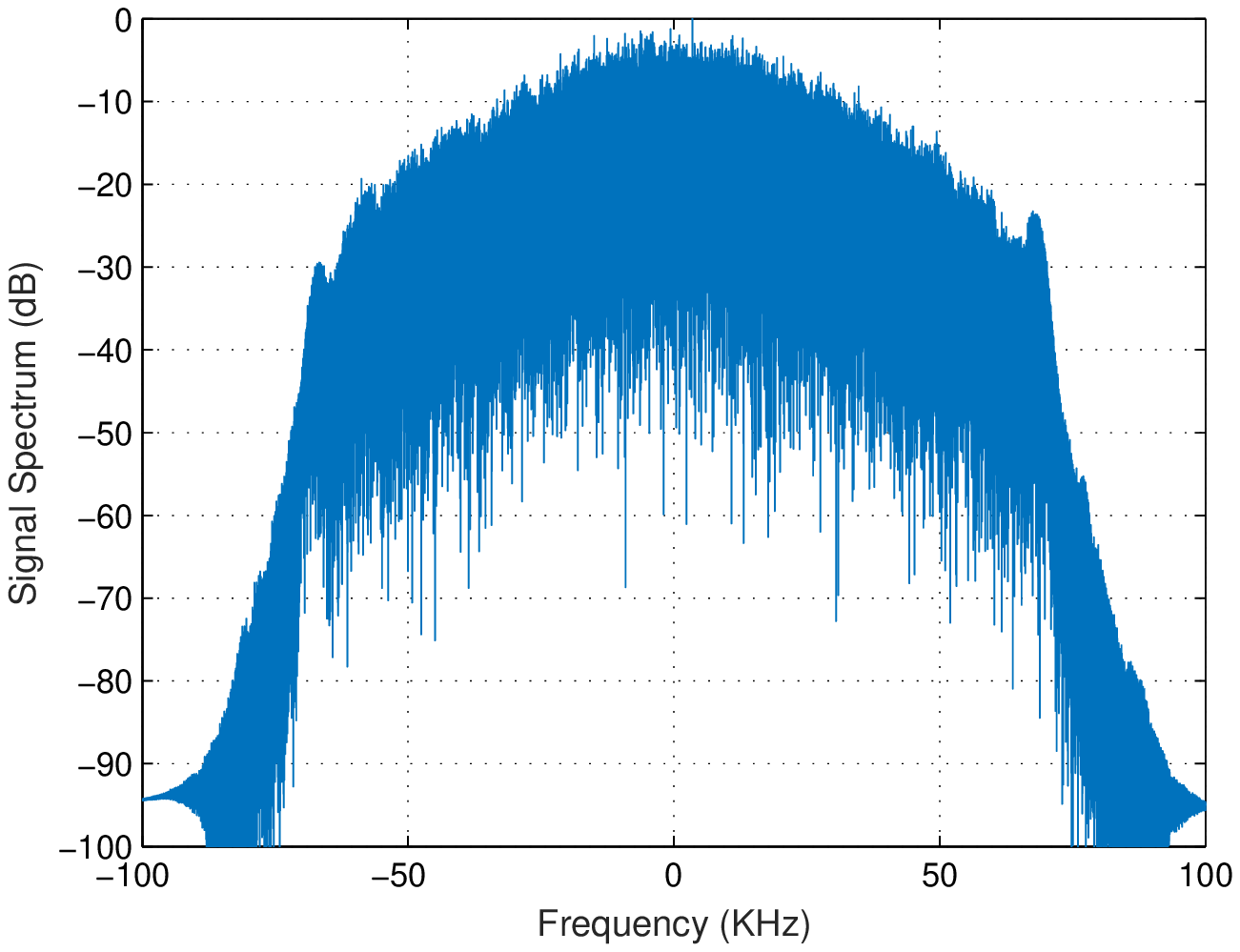}}
  \centering
 \caption{Normalized spectra of (a) modulating signal and (b) FM signal used for simulated scenario.} \label{Fig:FMsignal}
  \end{figure}
Then the conventional beamforming method can be used to estimate the angle of targets by searching the maximum of spectrum function
\begin{equation}\label{bm}
\mathbf{P}_m(\theta )=\mathbf{a}^H(\theta )\mathbf{z_mz_m}^H\mathbf{a}(\theta ), m=1,\cdots, N_T.
\end{equation}

\emph{Remark} 2: The angle of target can be estimated at first. Thus the estimation accuracy of angle can be improved, since no accumulative error of range-Doppler estimation can affect the estimation accuracy of angle. However, this method has a limitation, that the number of targets cannot exceed the number of antennas. Our proposed method does not have this limitation, the trade-off is that the estimation accuracy of angle in the proposed algorithm is lower than that of the method which estimates angle of the target at first.

\vspace{-1em}
\section{Simulation results}
\vspace{-1em}
\subsection{Modeling of the Transmitted Signal}
The modeling of transmitted signal is very important, since it would affect the performance of disturbance cancellation and parameter estimation. The most common signals for PBRs in use today are noncooperative FM commercial radio stations since they offer a good trade-off between performance and the system cost \cite{Colone2009multistage}. In order to simulate the FM radio signal, we take a segment of a song and then do frequency modulation such that the modulated signal has a bandwidth of about $100$kHz. The normalized spectra of modulating signal and FM signal used for simulated scenario are depicted in Fig. \ref{Fig:FMsignal}. It should be noted that the spectrum of the FM signal is similar to the one used in \cite{Colone2009multistage}, which verifies the correctness of the modeling of the transmitted signal.

Then we will evaluate the self-ambiguity property of the FM signal in the range-Doppler domain. The delay-Doppler auto-ambiguity function (AAF) of signal $s(t)$ over a time window $[0,T_0)$ can be written as
\begin{equation}\label{AAF}
\xi_0(\tau, f_D)=\int_{0}^{T_0}s(t)s^*(t-\tau)e^{-j2{\pi}f_Dt}dt.
\end{equation}

It can be seen from (\ref{AAF}) that the AAF measures the ambiguity level of a signal subject to a time delay $\tau$ and a Doppler shift $f_D$.  We plot in Fig. \ref{Fig:2DAAF} the 2D AAF of the FM signal presented in Fig. \ref{Fig:FMsignal} in the range-Doppler plane as well as its zero range cut and zero Doppler cut in Fig. \ref{Fig:RD}, where $T_0=1$s.
The strongest sidelobes appear at around $\pm65$Hz with small range. A further study shows that the peak-to-sidelobes ratio (PSLR) of the AAF is about $20$dB which is similar to the practical scenario \cite{Colone2009multistage}.

\begin{figure}
  \centering
  \subfigure[]{
    \label{Fig:2DAAF_0range}
    \includegraphics[width=1.6in,height=1in]{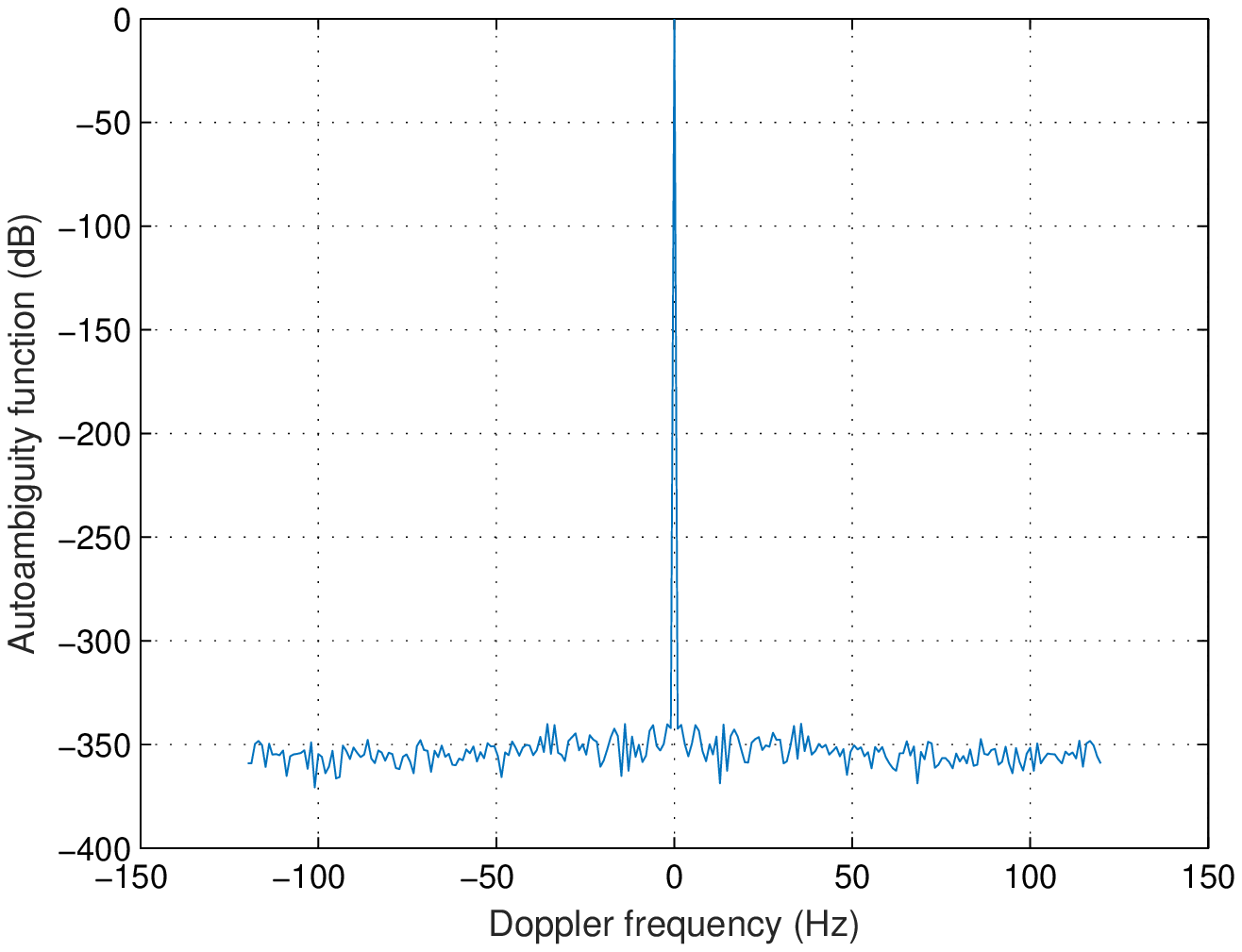}} %
  \subfigure[]{
    \label{Fig:2DAAF_0Doppler}
    \includegraphics[width=1.6in,height=1in]{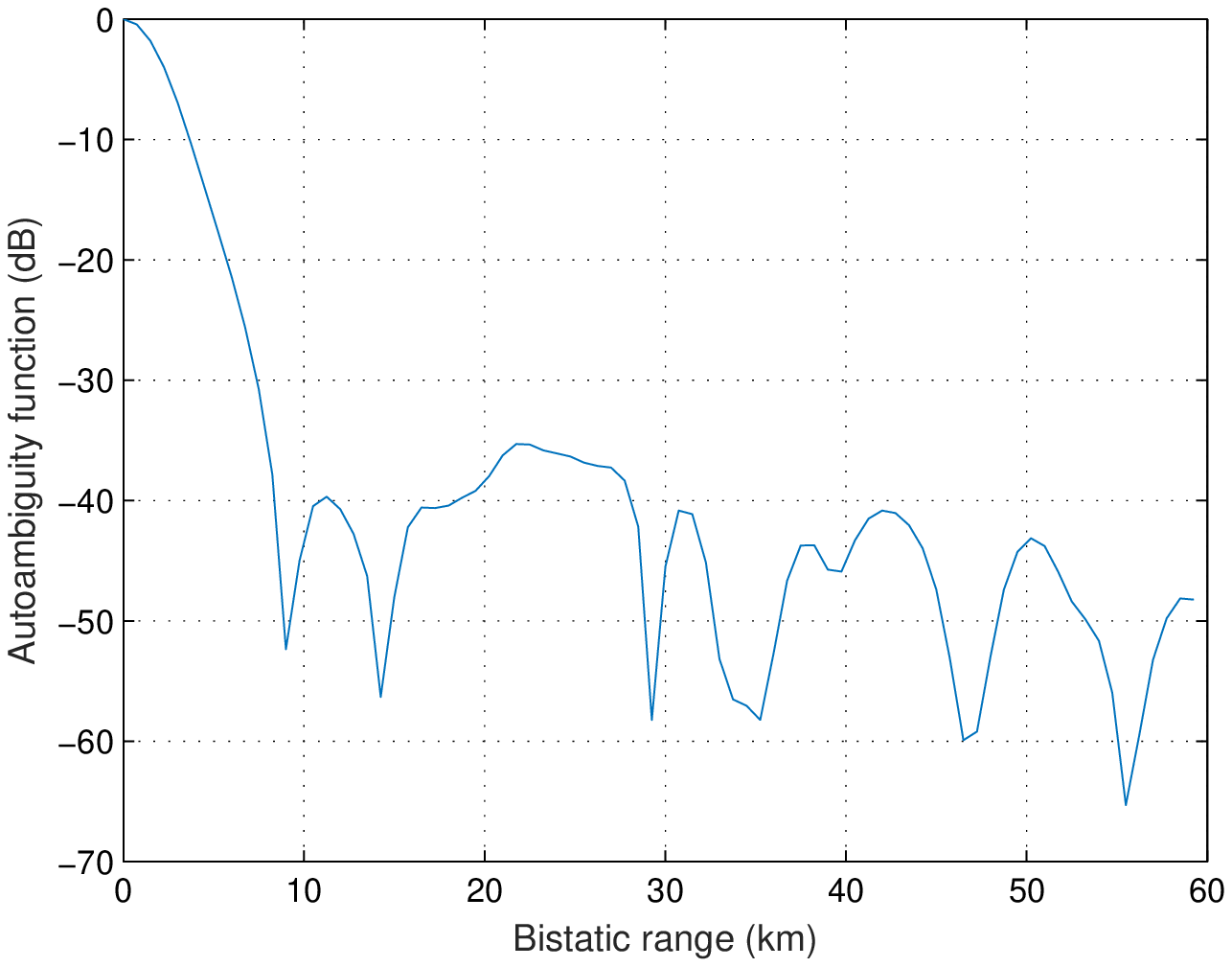}}
  \centering
  \caption{Normalized 2D AAF of FM signal used for simulated scenario. (a) Zero range cut. (b) Zero Doppler cut.} \label{Fig:AAF}
  \end{figure}

\vspace{-1em}
\subsection{Numerical Simulation for Parameter Estimation}
\begin{figure}
  \subfigure[]{
    \label{Fig:RDP}
    \includegraphics[width=1.6in,height=1in]{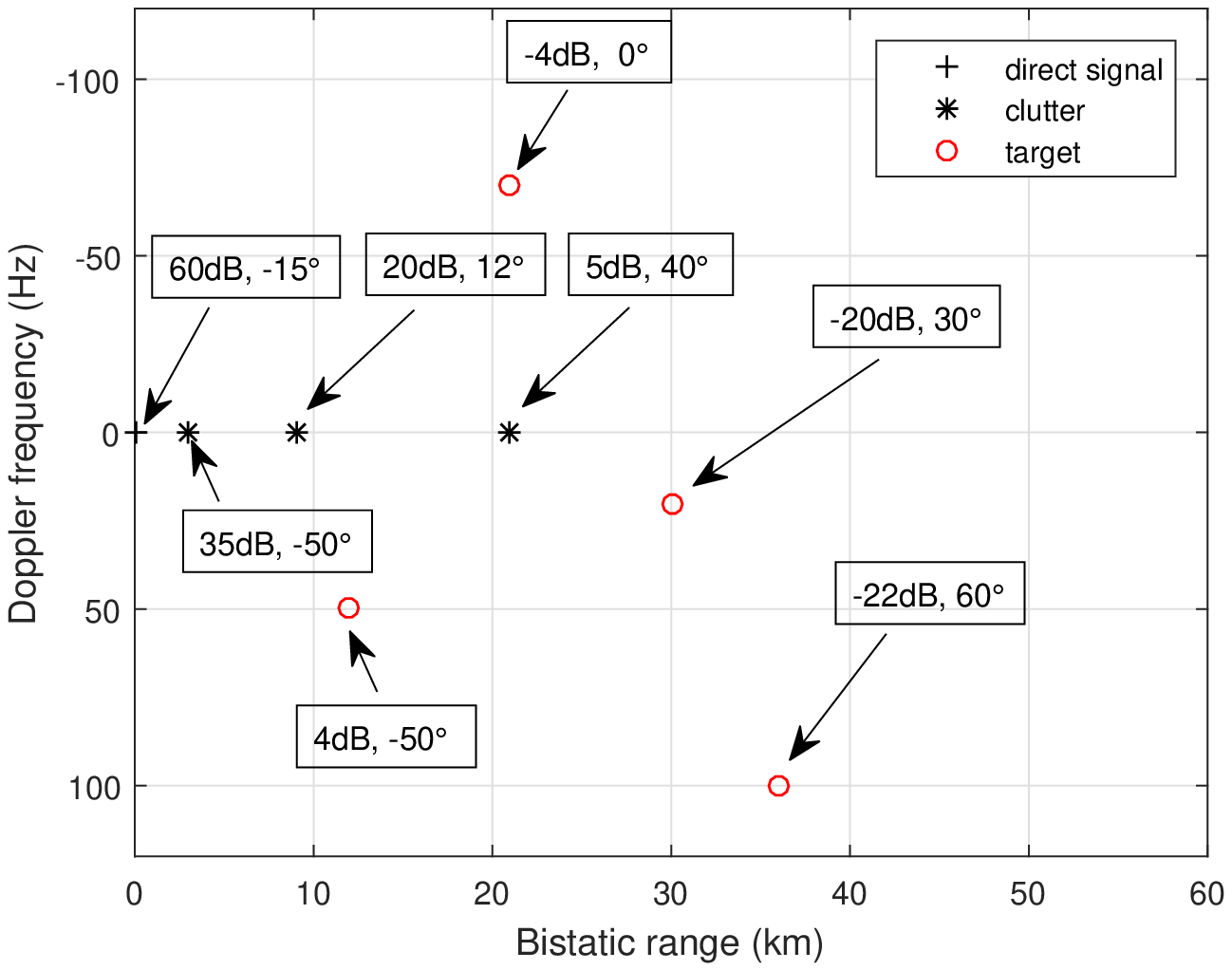}} %
  \subfigure[]{
\label{Fig:AP}
    \includegraphics[width=1.6in,height=1in]{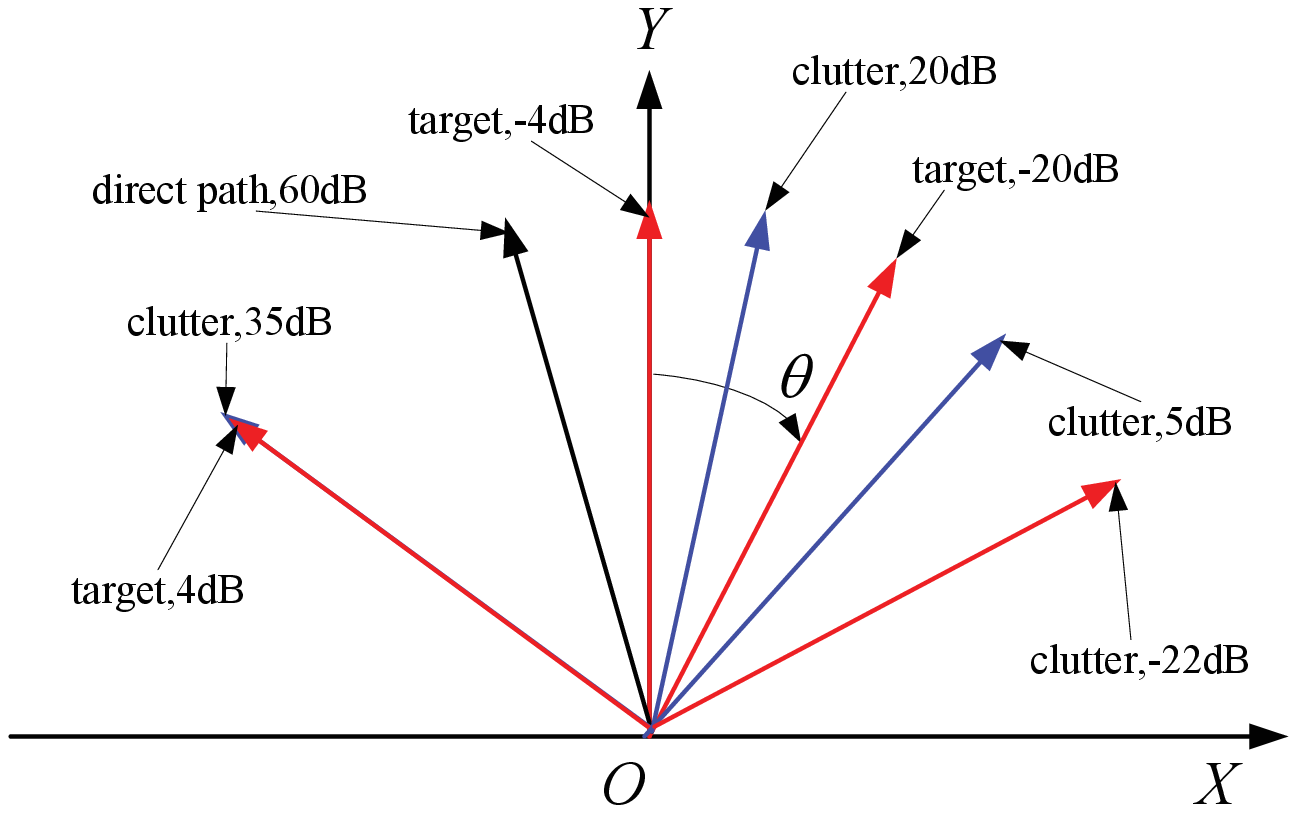}}
  \centering
  \caption{Sketch of the reference scenario. (a) In the range-Doppler plane. (b) In the angle plane. $\theta$ is positive when $OY$ turn in a clockwise; otherwise, negative.} \label{Fig:RD}
  \end{figure}

\begin{figure}
  \subfigure[]{
    \label{Fig:ECA}
    \includegraphics[width=1.6in,height=1in]{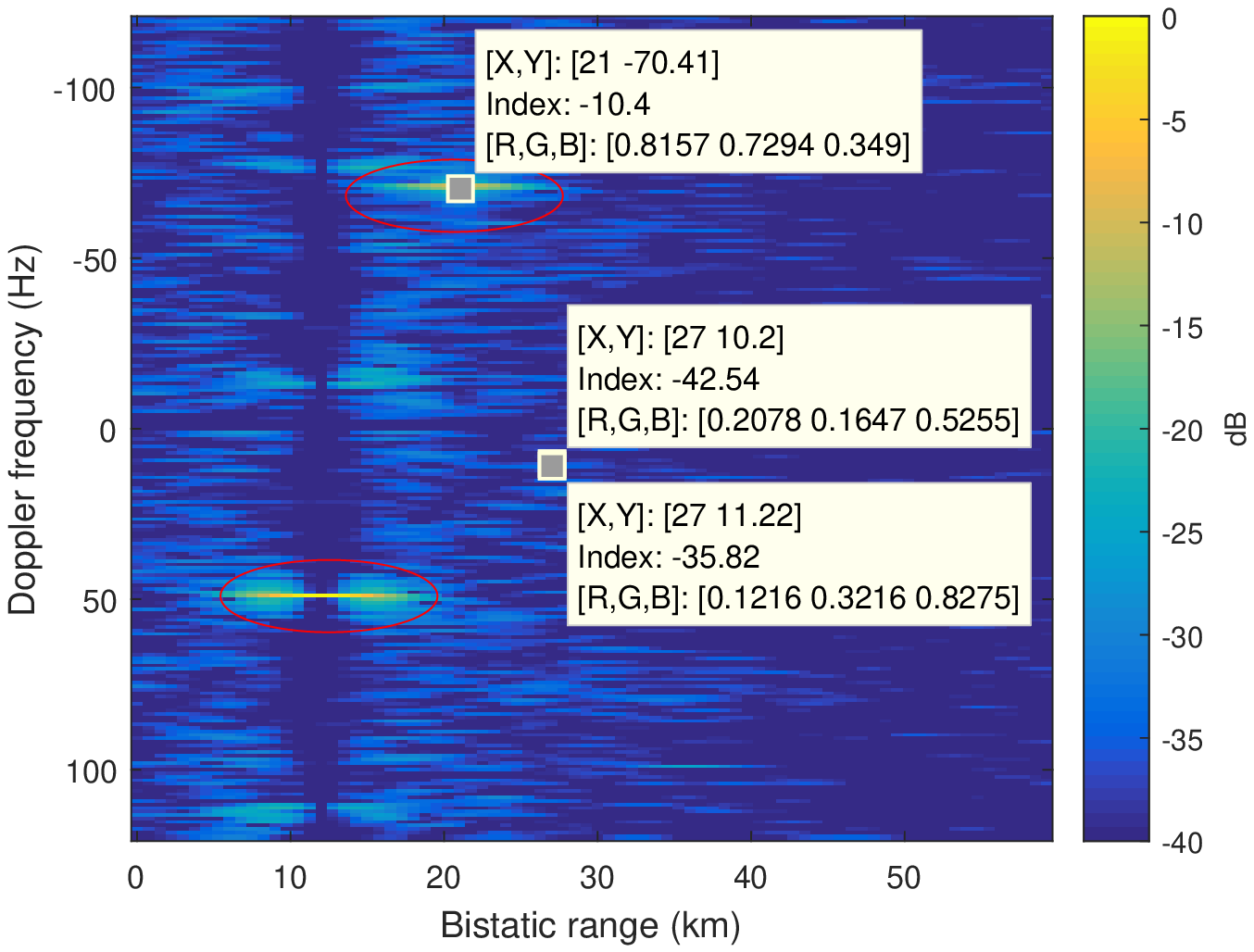}} %
  \subfigure[]{
\label{Fig:MECA}
    \includegraphics[width=1.6in,height=1in]{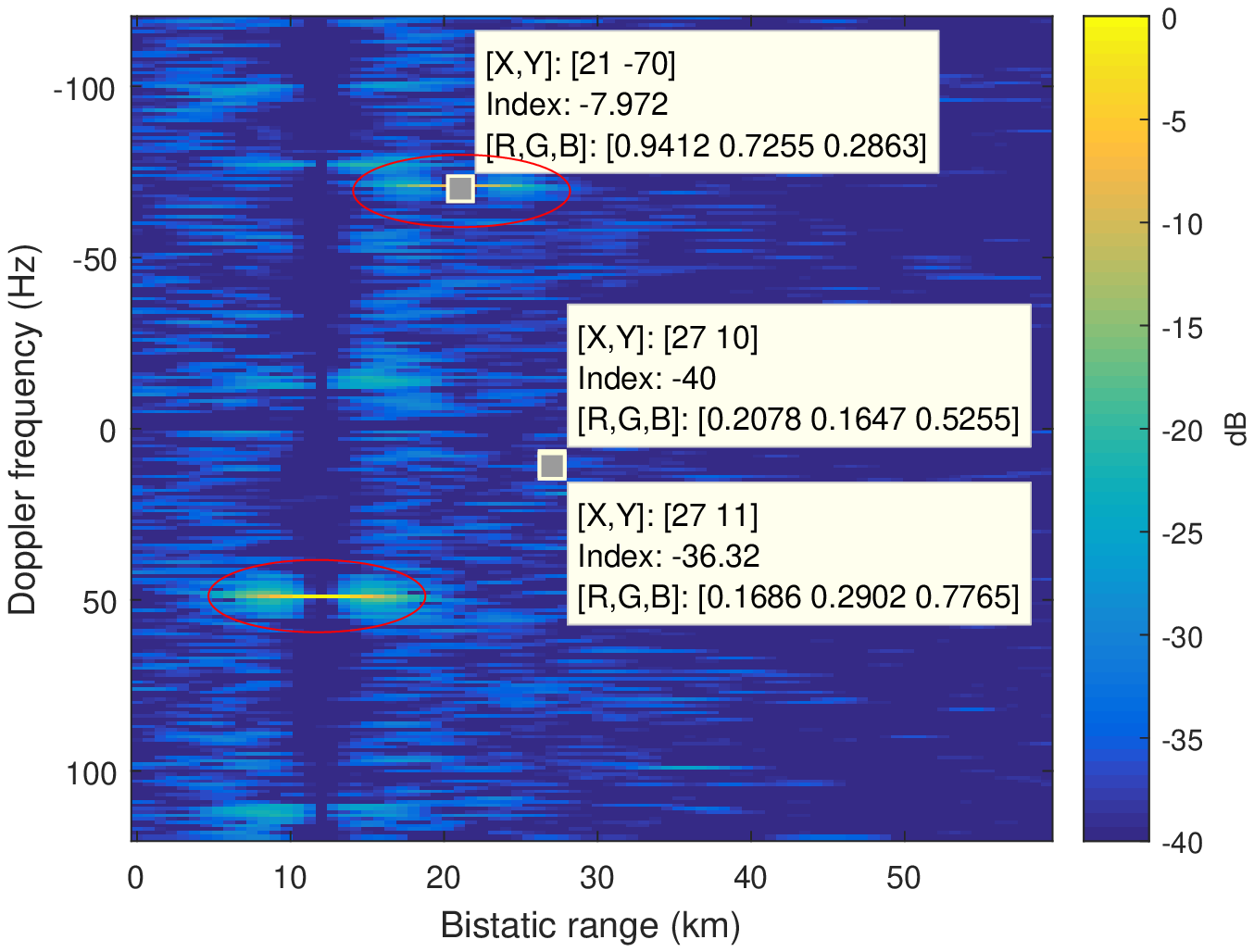}}
  \centering
  \caption{Normalized 2D CCF with (a) ECA and (b) MECA.}
  \end{figure}

We focus on range-Doppler-angle estimation. An example of the range-Doppler and angle domains are given in Fig. \ref{Fig:RDP} and Fig. \ref{Fig:AP}, respectively. The sampling rate is $400$kHz, i.e., four times of the bandwidth of the transmitted signal. The time window for parameter estimation is fixed at $1$s, which means the duration of the received array data that we can use must be shorter than 1s for ECA. Here it is fixed at $0.99$s.
First, the ECA method in \cite{Colone2009multistage} of disturbance cancellation is compared with our proposed method MECA.  It can be seen from Fig. \ref{Fig:ECA} that the spacing between two adjacent Doppler frequencies is not an integer, thus the sidelobes of clutters cannot be cancelled completely. However, from Fig. \ref{Fig:MECA}, it can be seen that the spacing between two adjacent Doppler frequencies is an integer, the MECA method is corresponding to integer Doppler frequencies, thus the sidelobes of clutters can be cancelled completely. In addition, it can be seen that the power of 2D-CCF of the second strong target in Fig. \ref{Fig:MECA} is stronger than that in Fig. \ref{Fig:ECA}, which verifies the reason mentioned above and the superiority of our proposed MECA method.

\begin{figure}
   \subfigure[]{
   \label{Fig:2DAAF}
    \includegraphics[width=1.6in,height=1in]{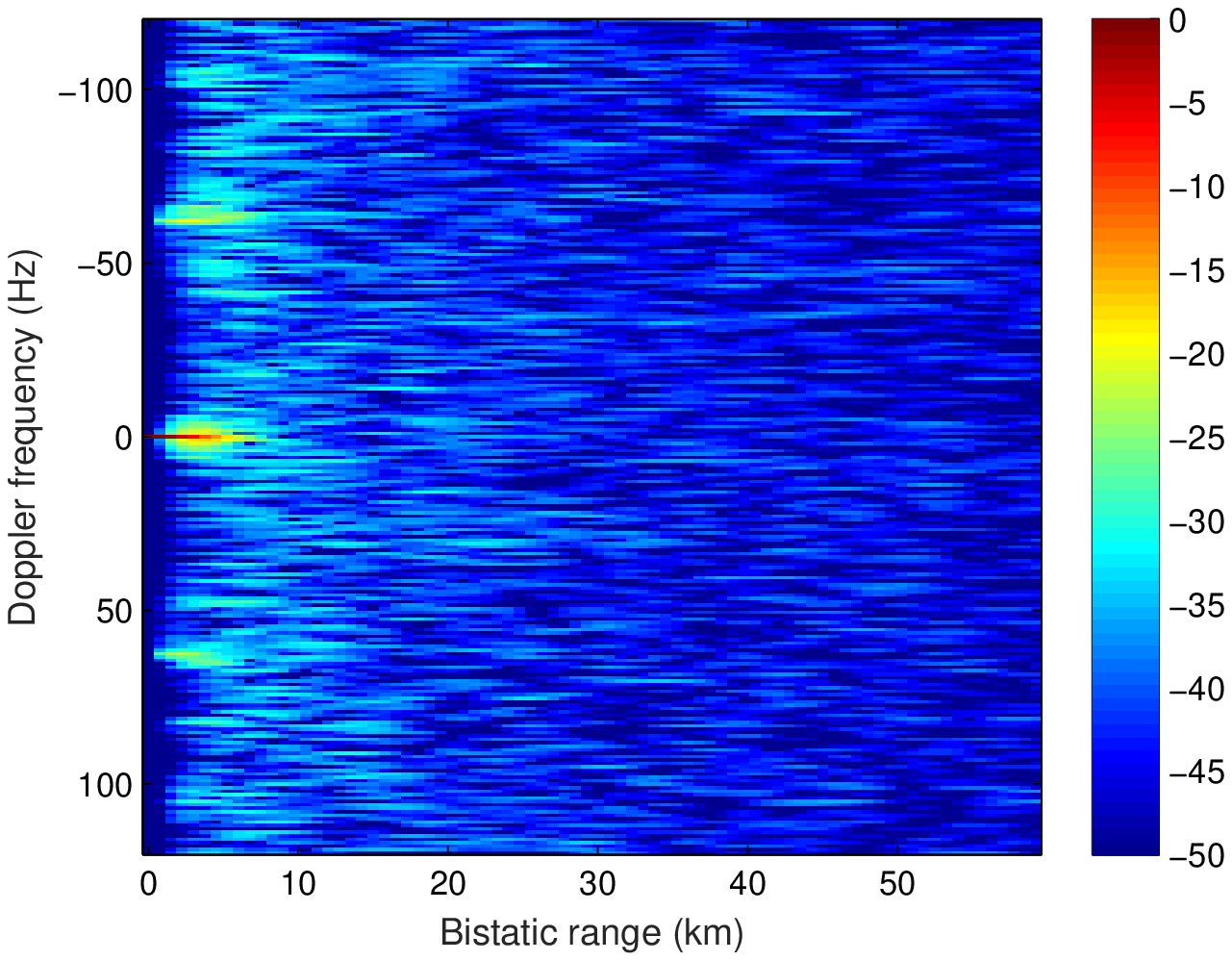}} %
  \subfigure[]{
  \label{Fig:WT}
    \includegraphics[width=1.6in,height=1in]{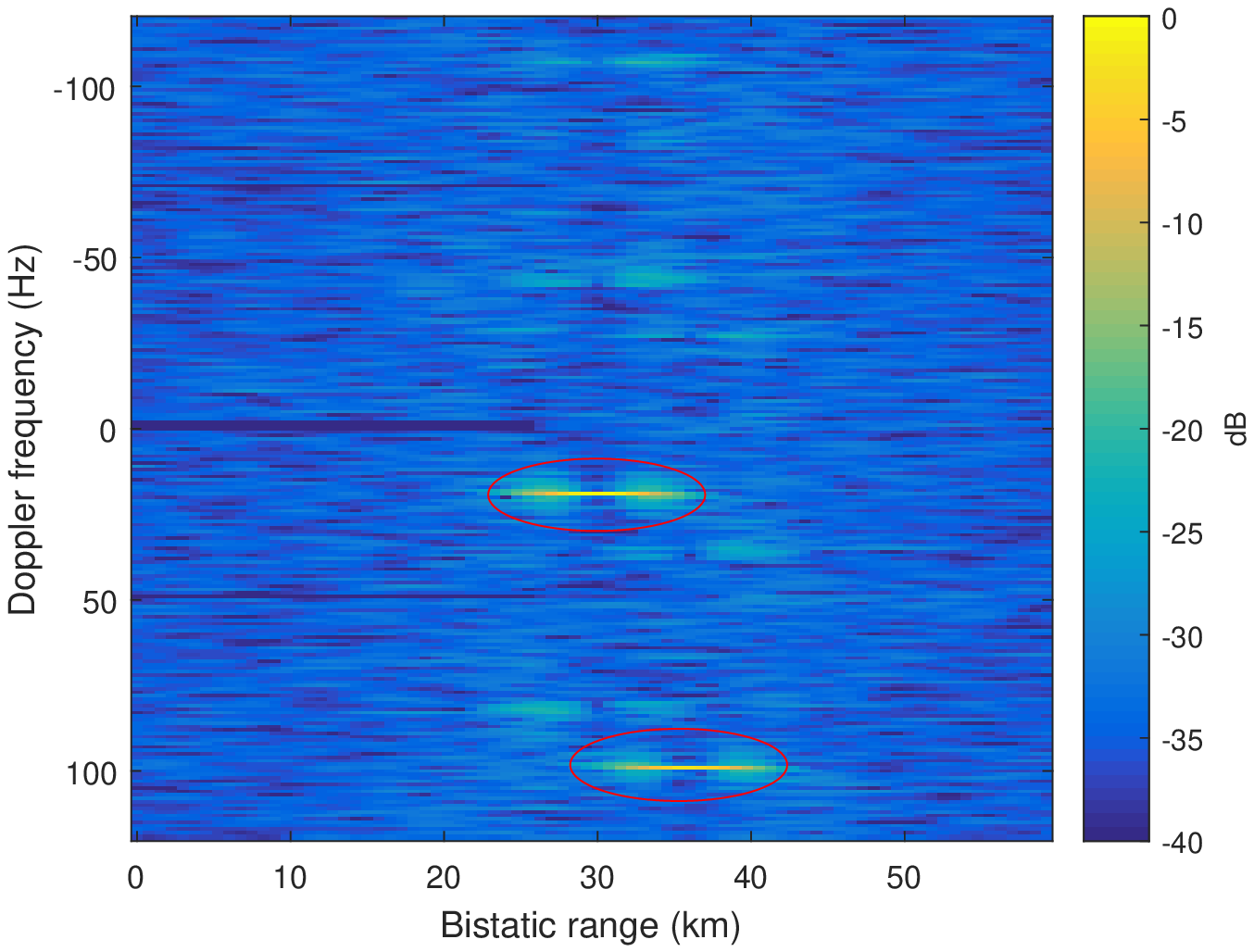}}
\caption{(a)Normalized 2D AAF of FM signal used for simulated scenario in 2D representation on range-Doppler plane. (b) Normalized 2D CCF with weak targets}
\end{figure}
\begin{figure}
  \subfigure[]{
    \includegraphics[width=1.6in,height=0.9in]{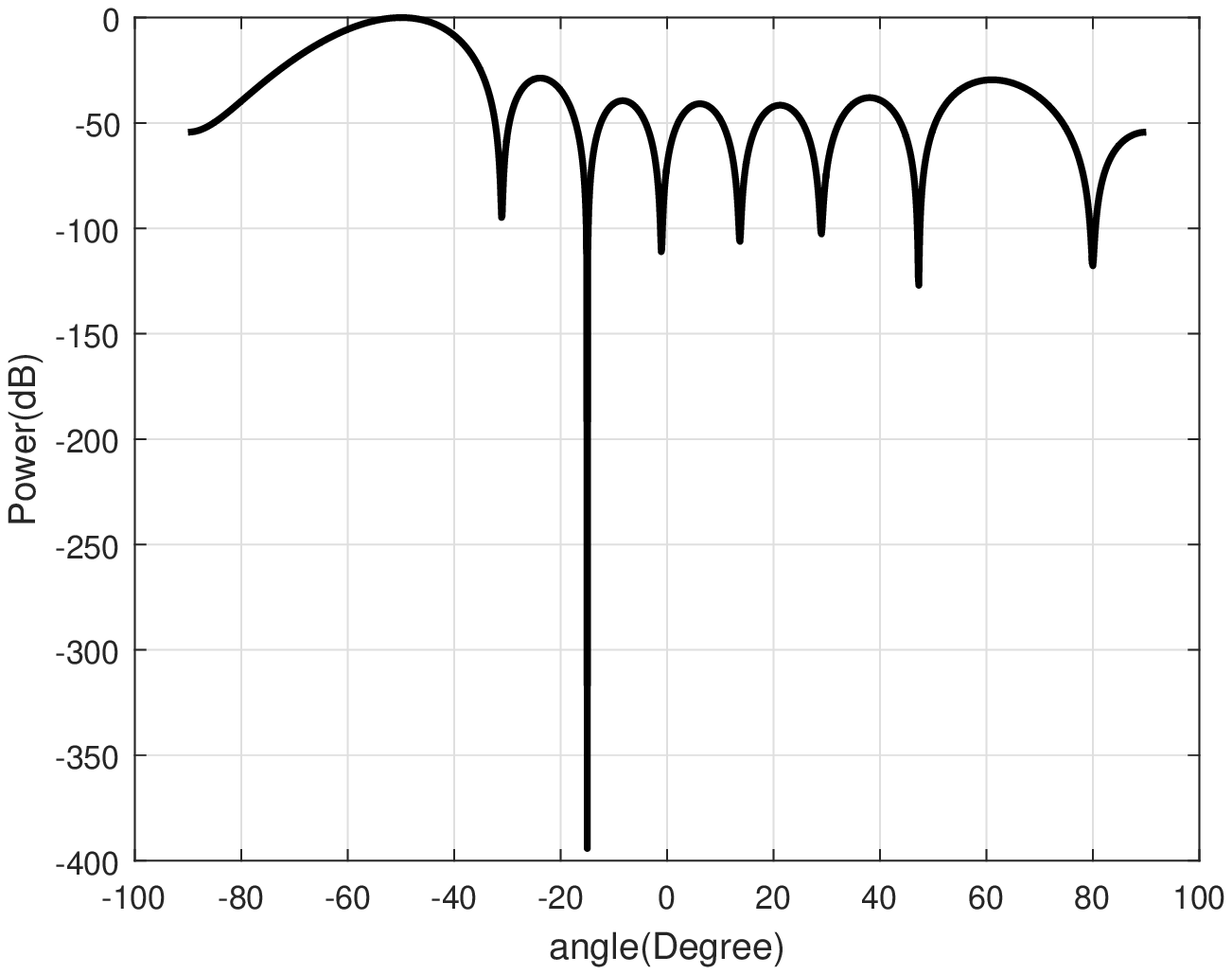}} %
  \subfigure[]{
    \includegraphics[width=1.6in,height=0.9in]{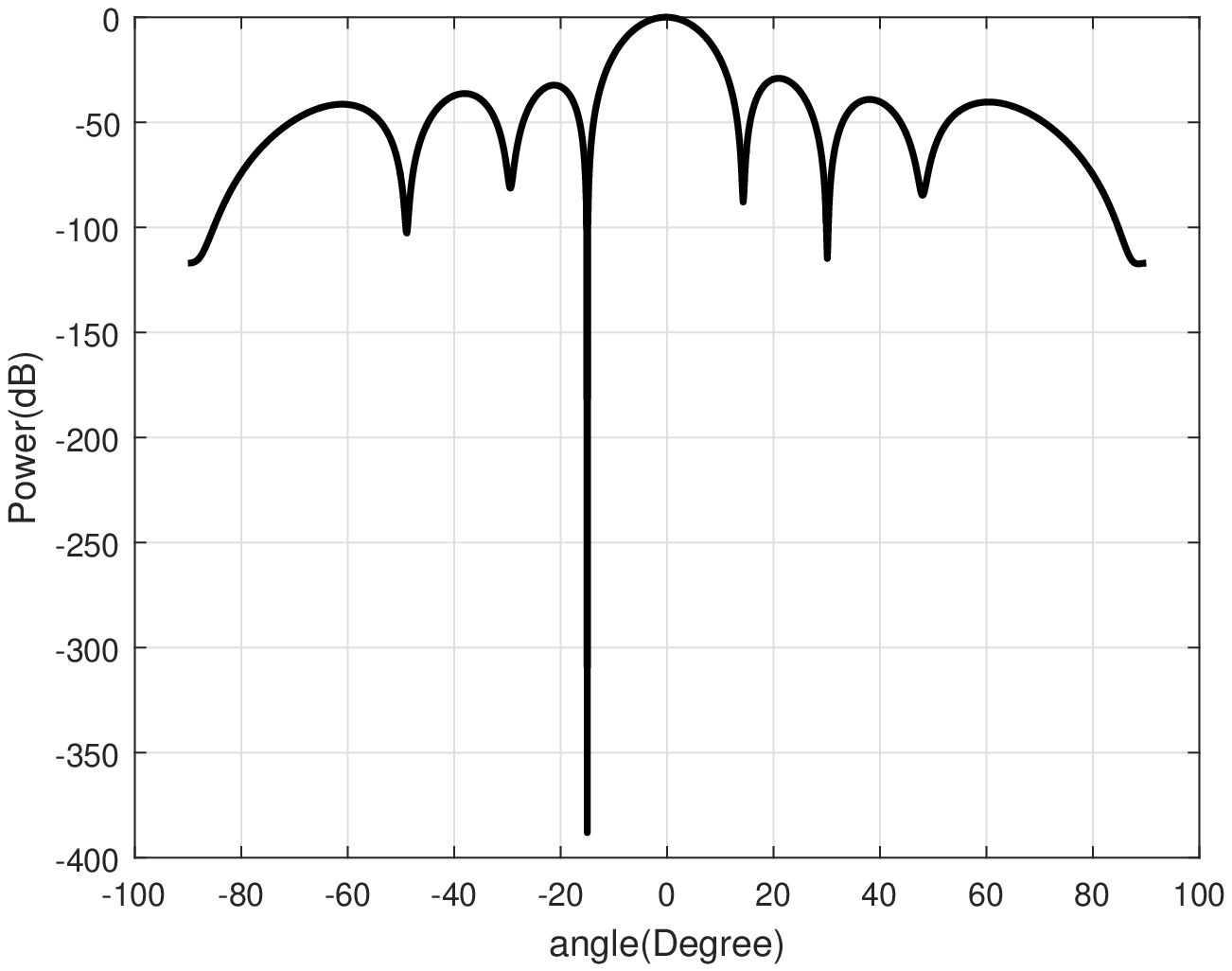}}
  \subfigure[]{
    \includegraphics[width=1.6in,height=0.9in]{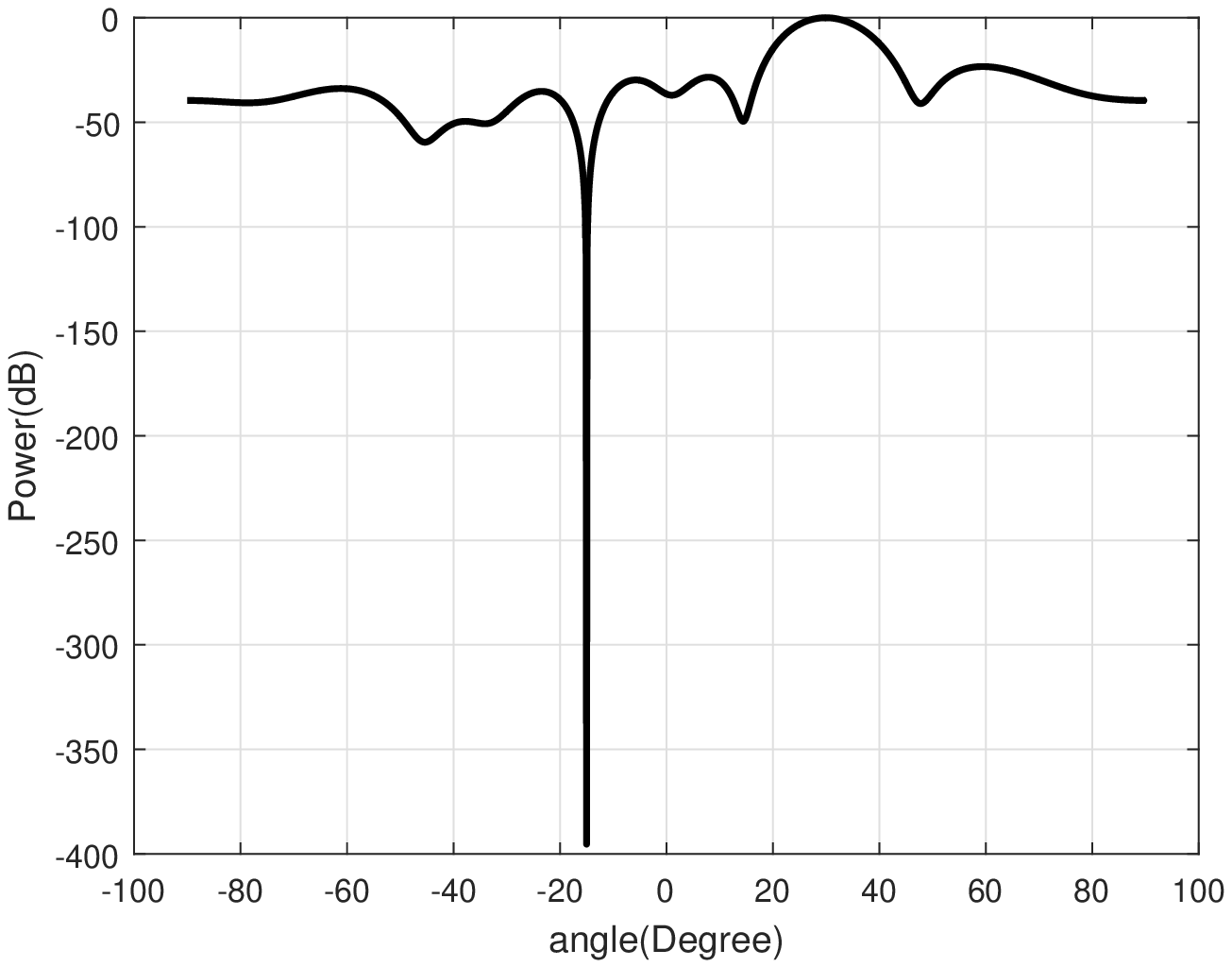}} %
  \subfigure[]{
    \includegraphics[width=1.6in,height=0.9in]{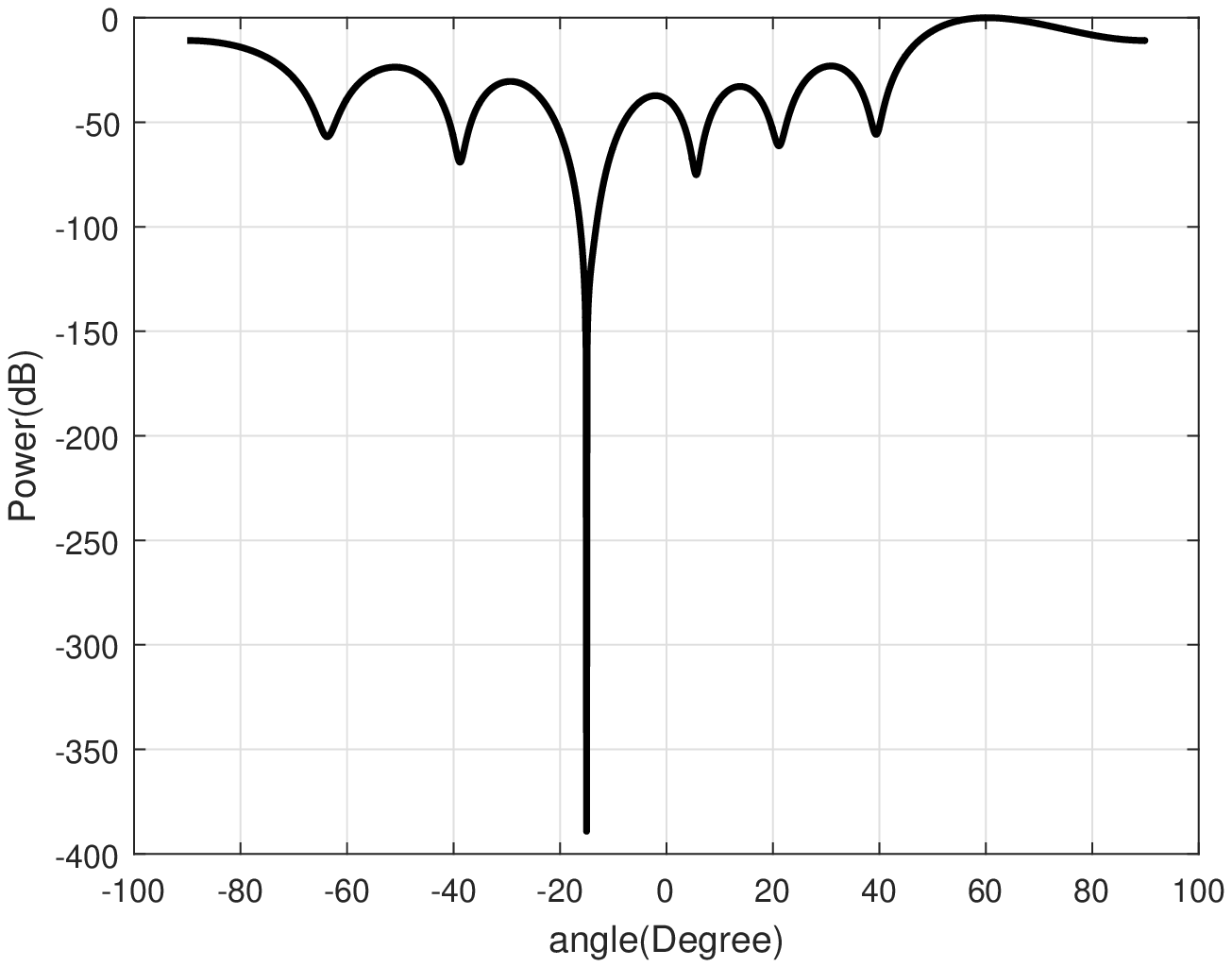}}
  \centering
  \caption{Angle estimation results using beamforming.}
\label{Fig:Ang}
  \end{figure}
Next, the range-Doppler-angle are estimated after using MECA method. It can be seen from Fig. \ref{Fig:MECA} that two strong targets appear in the RDmap. Then they should be cancelled based on MECA method to detect weak targets. It can be seen from Fig. \ref{Fig:WT} that the weak targets appear after cancelling the strong targets. Thus the range-Doppler estimation of the target of interest have been completed. Finally, the angle of the target of interest is estimated based on beamforming method as shown in Fig. \ref{Fig:Ang} .

\vspace{-1em}
\section{Conclusions}
In this paper, a direct path-range-Doppler-angle estimation method is proposed in PBR system. The direct path is estimated based on beamforming method. Then the range-Doppler of targets are estimated by using MF and LS methods after the disturbance cancellation based on MECA. The angles of targets are estimated by using beamforming method. The proposed estimation method is relatively simple, and is suitable for parameter estimation in real application.


\section{Acknowledgment}
The author would thank Dr. Danny Kai Pin Tan for his useful comments on disturbance cancellation.

\vfill\pagebreak

\bibliographystyle{IEEEbib}

\bibliography{reference}

\end{document}